\documentclass[]{elsarticle}

\usepackage{graphicx}
\usepackage{epsfig}		
\usepackage{dcolumn}
\usepackage{bm}
\usepackage{amsmath}
\usepackage{braket}
\usepackage{natbib}
\usepackage{color}
\usepackage{tikz-feynman}
\usepackage{feynmf}
\usepackage{pifont}
\usepackage{geometry}
\usepackage{fleqn}
\usepackage{txfonts}
\usepackage{amssymb }

\biboptions{numbers,sort&compress}

\def\0{\mbox{\tiny $0$}}
\def\1{\mbox{\tiny $1$}}
\def\2{\mbox{\tiny $2$}}
\def\3{\mbox{\tiny $3$}}
\def\4{\mbox{\tiny $4$}}
\def\5{\mbox{\tiny $5$}}
\def\6{\mbox{\tiny $6$}}
\def\7{\mbox{\tiny $7$}}
\def\8{\mbox{\tiny $8$}}
\def\9{\mbox{\tiny $9$}}

\def\f14{\mbox{\tiny $\frac{1}{4}$}}

\journal{Physics Letters B}

\begin{document}


\title{Scalar field dark matter with spontaneous symmetry breaking and the $3.5$ keV line}

\author[ob]{Catarina Cosme}
\ead{catarinacosme@fc.up.pt}
\author[jr]{Jo\~{a}o G. Rosa}
\ead{joao.rosa@ua.pt}
\author[ob]{O. Bertolami}
\ead{orfeu.bertolami@fc.up.pt}

\address[ob]{Departamento de F\'isica e Astronomia, Faculdade de Ci\^{e}ncias da
Universidade do Porto and  Centro de F\'isica do Porto, Rua do Campo Alegre 687, 4169-007, Porto, Portugal.}
\address[jr]{Departamento de F\'isica da Universidade de Aveiro and Center for Research and Development in Mathematics and Applications (CIDMA) Campus de Santiago, 3810-183 Aveiro, Portugal.}

 
\date{\today}

\begin{abstract}

We show that the present dark matter abundance can be accounted for by an oscillating scalar field that acquires both mass and a non-zero expectation value from interactions with the Higgs field. The dark matter scalar field can be sufficiently heavy during inflation, due to a non-minimal coupling to gravity, so as to avoid the generation of large isocurvature modes in the CMB anisotropies spectrum. The field begins oscillating after reheating, behaving as radiation until the electroweak phase transition and afterwards as non-relativistic matter. The scalar field becomes unstable, although sufficiently long-lived to account for dark matter, due to mass mixing with the Higgs boson, decaying mainly into photon pairs for masses below the MeV scale. In particular, for a mass of $\sim 7$ keV, which is effectively the only free parameter, the model predicts a dark matter lifetime compatible with the recent galactic and extragalactic observations of a 3.5 keV X-ray line.

\end{abstract}


\maketitle



One of the most important open problems in modern cosmology is the nature of dark matter (DM), an invisible form of matter that can explain the observed structure of the Universe on large scales, the galaxy rotation curves and the anisotropies in the Cosmic Microwave Background (CMB). However, despite the large number of candidates, there are still no definite answers concerning its origin \cite{Bertone:2004pz}. An interesting possibility is an interaction between DM and the Higgs field, widely known as ``Higgs-portal DM''. This has been extensively studied in the literature, namely in the context of thermal production \cite{Patt:2006fw,Bento:2001yk,MarchRussell:2008yu,Biswas:2011td,Pospelov:2011yp,Mahajan:2012nc,Cline:2013gha,Enqvist:2014zqa,Kouvaris:2014uoa,Costa:2014qga,Duerr:2015aka,Han:2015dua,Han:2015hda}. However, the lack of evidence for WIMP-like particles in the various ongoing experiments \cite{Baer:2014eja} suggests looking for alternative candidates, such as oscillating scalar fields, as considered e.g.~in Refs. \cite{Tenkanen:2015nxa,Kainulainen:2016vzv,Bertolami:2016ywc}.

In this Letter, we show for the first time that a scalar field dark matter coupled to the Higgs field can naturally explain the 3.5 keV X-ray line detected by the XMM-Newton observatory. Our model considers a complex scalar field, $\Phi$, interacting with the Higgs doublet, $\mathcal{H}$, only through scale-invariant interactions given by the Lagrangian density:
\begin{equation}
\mathcal{L}_{int}=\pm\, g^{2}\left|\Phi\right|^{2}\left|\mathcal{H}\right|^{2}+\lambda_{\phi}\left|\Phi\right|^{4}+V\left(\mathcal{H}\right)+\xi R\left|\Phi\right|^{2}~,\label{Lagrangian}
\end{equation}
where the Higgs potential $V(\mathcal{H})$ has the standard ``mexican hat" shape.  We assume that the scale invariance of the $\Phi$ interactions is a consequence of an underlying scale invariance of the full theory, that is spontaneously broken in the Higgs and gravitational sectors by some mechanism that has no influence on the effective dynamics of the dark matter scalar field (see also Ref. \cite{Heikinheimo:2017ofk}). This allows for the Higgs-dark scalar interaction with coupling, $g$, the dark scalar field quartic self-interactions with coupling, $\lambda_\phi$, and for a non-minimal coupling, $\xi$, of the DM to the Ricci scalar, $R$.

The interaction Lagrangian (\ref{Lagrangian}) also exhibits a U(1) symmetry and we may consider two cases. When the Higgs-dark scalar interaction has a positive sign, the U(1) symmetry remains unbroken and the DM field is stable. For a negative coupling, the U(1) symmetry can be spontaneously broken in the vacuum and the DM field may decay, allowing for astrophysical signatures, as we will see below. In this Letter, we focus on the latter case, leaving the discussion of the former to a longer companion paper.

The background dynamics of the homogeneous dark scalar field mode is determined by the equation of motion:
\begin{equation}
\ddot{\phi}+3H\dot{\phi}+V'\left(\phi\right)+2\xi R\phi=0~,\label{eom NMC}
\end{equation}
where $\Phi=\phi/\sqrt{2}$ since the complex phase has a trivial dynamics. From the associated energy-momentum tensor, we obtain the effective energy density and pressure of the field, which are, respectively, given by 
\begin{equation}
\rho_{\phi}=\frac{\dot{\phi}^{2}}{2}+V\left(\phi\right)+12\xi H\phi\dot{\phi}+6\xi\phi^{2}H^{2}~,\label{energy density}
\end{equation}
\begin{equation}
p_{\phi}=\frac{1}{2}\,\left(1-8\xi\right)\dot{\phi}^{2}-V\left(\phi\right)+4\xi\phi V'\left(\phi\right)+4\xi\phi\dot{\phi}H+\xi\phi^{2}\left[\left(8\xi-1\right)R+2\,\frac{\ddot{a}}{a}+4\,H^{2}\right]~,\label{pressure}
\end{equation}
where $a$ is the scale factor. We will see below that the introduction of a non-minimal coupling does not significantly change the usual form of $\rho_{\phi}$ and $p_{\phi}$ for an oscillating scalar field. As pointed out in Ref. \cite{Bertolami:2016ywc}, the initial conditions for the scalar field oscillations are set by the inflationary dynamics. In the parametric regime where $\xi\gg g,\lambda_\phi$, which will henceforth be the focus of our discussion, the field's mass during inflation is dominated by the non-minimal coupling to the curvature scalar,  $R\simeq 12\,H_{inf}^{2}$, where $H_{inf}\simeq2.5\times10^{13}\left(r/0.01\right)^{1/2}$ GeV is the Hubble parameter during inflation and $r$ is the tensor-to-scalar ratio. This yields  $m_{\phi}\simeq\sqrt{12\xi}\,H_{inf}\gtrsim H_{inf}$ for $\xi\gtrsim 0.1$. As pointed out in Ref. \cite{Bertolami:2016ywc}, this super-Hubble mass prevents the field from acquiring large fluctuations during inflation that would give rise to observable isocurvature modes in the CMB spectrum, which are now significantly constrained \cite{Ade:2015lrj}. For  $m_{\phi}/H_{inf}>3/2$, quantum fluctuations in the field get stretched and amplified during inflation, yielding a spectrum \cite{Riotto:2002yw}:
\begin{equation}
\left|\delta\phi_{k}\right|^{2}\simeq\left(\frac{H_{inf}}{2\pi}\right)^{2}\left(\frac{H_{inf}}{m_{\phi}}\right)\frac{2\pi^{2}}{\left(a\,H_{inf}\right)^{3}}~.\label{Fourier modes}
\end{equation}
Integrating over the comoving momentum $k$ on super-horizon scales, we can obtain the field variance at the end of inflation, which sets the typical homogeneous field amplitude at the onset of oscillations in the post-inflationary era, $\phi_{inf}$:
\begin{equation}
\phi_{inf}=\sqrt{\left\langle \phi^{2}\right\rangle }\simeq\alpha\,H_{inf},\qquad\alpha\simeq0.05\,\xi^{-1/4}~.\label{eq: initial amplitude after inflation}
\end{equation}
Note that, during inflation, all terms in Eq.~(\ref{energy density}) are $\sim H_{inf}^{4}$ and therefore the dark scalar plays a negligible role in the inflationary dynamics.

We should briefly mention that, during the (p)reheating period, the Ricci scalar oscillates with the inflaton field, $\chi$, since $R=(3p_\chi-\rho_\chi)/M_{Pl}^2\sim m_\chi^2\chi^2/M_{Pl}^2$, inducing an effective biquadratic coupling between the dark scalar and the inflaton, $g_{\phi\chi}^2\sim \xi m_\chi^2/M_{Pl}^2\ll 1$. This interaction will lead to $\phi$-particle production during reheating but, since $q_\phi=g_{\phi\chi}^2\chi^2/4m_\chi^2\sim \xi \chi^2/M_{Pl}^2 \lesssim 1$ with $\chi\lesssim M_{Pl}$ during reheating, this should not be very efficient. In particular, it is natural to assume that the inflaton couples more strongly to other fields, which will thus be produced more efficiently and consequently reduce the amplitude of the inflaton's oscillations before any significant $\phi$-particle production occurs. In addition, such particles remain relativistic until $T< m_\phi \ll T_{EW}$, and as we will see this implies that their density is much more diluted than the density of the homogeneous dark scalar condensate. We therefore expect $\phi$-particle production during reheating to yield a negligible contribution to the present dark matter abundance.

After inflation and the reheating period, the Universe becomes dominated by radiation, and $R\simeq 0$. For temperatures above the electroweak scale, thermal effects keep the Higgs close to the origin (see e.g. \cite{Bastero-Gil:2016mrl}), such that the dark scalar field potential is dominated by the quartic term, $V\left(\phi\right)\simeq \lambda_{\phi}\phi^{4}/4$. Once the effective field mass $m_{\phi}=\sqrt{3\lambda_{\phi}}\phi$ exceeds the Hubble parameter in this era, the field starts oscillating about the origin with an amplitude that decays as $a^{-1}\propto T$. 

It is easy to check that, in the oscillating phase, the last two terms in Eqs. (\ref{energy density}) and (\ref{pressure}) become subdominant since $m_\phi\gg H$. In addition, the remaining terms in Eq.~(\ref{pressure}) proportional to $\xi$ cancel out upon averaging over the field oscillations, since  $\langle \dot{\phi}^{2}\rangle =\langle \phi\,\mathrm{V}'\left(\phi\right)\rangle $. This implies that the field's energy density and pressure are approximately given by the corresponding $\xi=0$ expressions once it begins oscillating, such that $\rho_\phi \propto a^{-4}$ as long as the quartic potential term is dominant. During this period, the field thus behaves as {\it dark radiation}.

Equating the Hubble parameter in the radiation era with the effective field mass, we obtain for the cosmic temperature at the onset of field oscillations:
\begin{equation}
T_{rad}=\left(\sqrt{3\lambda_{\phi}}\,\phi_{inf}\,M_{Pl}\,\sqrt{\frac{90}{\pi^{2}g_{*}}}\right)^{1/2}~,\label{T rad}
\end{equation}
where $g_{*}$ is the number of relativistic degrees of freedom. This is below the reheating temperature if the inflaton decays sufficiently fast after inflation, with $T_R\sim \sqrt{H_{inf} M_P}$ for instantaneous reheating.

Once the temperature drops below the electroweak scale, the Higgs field acquires a vacuum expectation value (vev) and the relevant Lagrangian density for the real $\phi$ and Higgs components is:
\begin{equation}
\mathcal{L}_{int}=-\frac{g^{2}}{4}\,\phi^{2}\,h^{2}+\frac{\lambda_{\phi}}{4}\,\phi^{4}+\frac{\lambda_{h}}{4}\left(h^{2}-\tilde{v}^{2}\right)^{2}~,\label{potential EWPT}
\end{equation}
where $\lambda_{h}\simeq0.13$ is the Higgs self-coupling. The Higgs and dark scalar vevs are, respectively:
\begin{equation}
h_{0}=\left(1-\frac{g^{4}}{4\,\lambda_{\phi}\,\lambda_{h}}\right)^{-1/2}\tilde{v}\equiv\mathrm{v},\quad\phi_{0}=\frac{g}{\sqrt{2\lambda_{\phi}}}\,\mathrm{v}~,\label{h0 as v}
\end{equation}
where $\mathrm{v}=246\,\mathrm{GeV}$. Note that a non-vanishing vev for $\phi$ implies $g^4< 4\lambda_\phi \lambda_h$, which we assume to hold.

The interaction Lagrangian above is valid once the leading thermal contributions to the Higgs potential become Boltzmann-suppressed, which should occur below $T_{EW}\sim m_W$, where $m_W$ is the $W$ boson's mass. At this point, the field starts oscillating about $\phi_0$ rather than about the origin. To determine the amplitude of oscillations at this stage, note that at $T_{EW}$ the amplitude of field oscillations about the origin has been redshifted to: 
\begin{equation}
\phi_{EW}\simeq\left(\frac{4\,\pi^{2}\,g_{*}}{270}\right)^{1/4}\,\left(\frac{\phi_{inf}}{M_{Pl}}\right)^{1/2}\,\frac{T_{EW}}{\mathrm{v}}\,\frac{\lambda_{\phi}^{1/4}}{g}\,\phi_{0}\nonumber
\simeq10^{-4}\,g_{*}^{1/4}\,\xi^{-1/8}\,\left(\frac{T_{EW}}{m_{W}}\right)\left(\frac{r}{0.01}\right)^{1/4}\,\frac{\lambda_{\phi}^{1/4}}{g}\,\phi_{0}.
\end{equation}
We thus see that $\phi_{EW}\lesssim \phi_0$ for $g\gtrsim 10^{-4}\lambda_\phi^{1/4}$ for $\xi\sim \mathcal{O}(1)$, with a larger non-minimal coupling to curvature localizing the field even closer to the origin at the electroweak phase transition (EWPT). This implies that, in these parametric regimes, the field will start oscillating about the non-zero vev below $T_{EW}$, with an amplitude $\phi_{DM}\equiv x_{DM} \,\phi_0$ with $x_{DM}\lesssim 1$ \cite{foot2}. The field's equation of state then smoothly changes from a dark radiation to a cold dark matter behavior as the potential becomes quadratic about the minimum.

Therefore, the field amplitude evolves with the temperature as $\phi\left(T\right)=\phi_{DM} (T/T_{EW})^{3/2}$ and the number of particles per comoving volume becomes constant: 
\begin{equation}
\frac{n_{\phi}}{s}=\frac{45}{4\pi^{2}g_{*S}}\frac{m_\phi\phi_{DM}^{2}}{T_{EW}^{3}}~,\label{n over s}
\end{equation}
where $g_{*S}\simeq 86.25$ is the number of relativistic degrees of freedom contributing to the entropy at $T_{EW}$. We can use this to compute the present DM abundance, $\Omega_{\phi,0}\simeq 0.26$, obtaining the relation:
\begin{equation}
m_{\phi}=\left(6\,\Omega_{\phi,0}\right)^{1/2}\left(\frac{g_{*S}}{g_{*S0}}\right)^{1/2}\left(\frac{T_{EW}}{T_{0}}\right)^{3/2}\frac{H_{0}M_{Pl}}{\phi_{DM}}~,\label{mass phi}
\end{equation}
where $H_{0}$, $g_{*S0}$ and $T_0$ are the present values of the Hubble parameter, number of relativistic degrees of freedom and CMB temperature, respectively. Given that $m_\phi= g\mathrm{v}$, this leads to the following relation between $g$ and $\lambda_{\phi}$:
\begin{equation}
g\simeq2\times10^{-3}\,\left(\frac{x_{DM}}{0.5}\right)^{-1/2}\,\lambda_{\phi}^{1/4}~.\label{g and lambd}
\end{equation}
Note that this consistently satisfies the parametric constraints for spontaneous symmetry breaking and $\phi_{EW}\lesssim \phi_0$ described above. This relation leaves essentially a single free parameter in the model, which we take to be the mass of the field.

There are, however, further constraints on this parameter that we must take into account. In particular, we have assumed that the scalar field remained as a homogeneous condensate throughout its whole evolution and that it never thermalized with the surrounding cosmic plasma. Otherwise, condensate evaporation would lead to a WIMP-like candidate for DM, the phenomenology of which was already studied in Ref. \cite{Enqvist:2014zqa}. There are two processes that lead to the evaporation of the condensate: the Higgs annihilation into higher-momentum $\phi$ particles and the perturbative production of $\phi$ particles by the oscillating background field.

Let us start by considering the Higgs annihilation, which for $T\gtrsim T_{EW}$ occurs at a rate:
\begin{equation}
\Gamma_{hh\rightarrow\phi\phi}=n_{h}\left\langle \sigma v\right\rangle~,\label{decay rate higgs}
\end{equation}
where $v\sim c\equiv1$ and $n_{h}$ is the Higgs number density. Before the EWPT, the typical momentum of Higgs particles  $\left|\vec{p}\right|\sim T$, so that the cross section of the process is given by:
\begin{equation}
\sigma\simeq\frac{g^{4}}{64\pi} T^{-2}\left(1+\frac{m_{h}^{2}}{T^{2}}\right)^{-1}\sqrt{1+\frac{m_{h}^{2}-m_{\phi}^{2}}{T^{2}}}~.\label{total cross section}
\end{equation}
After the EWPT, all Higgs bosons essentially decay into lighter Standard Model (SM) degrees of freedom and therefore $\phi$ production stops. Thus, to prevent the thermalization of $\phi$ particles we must require $\Gamma_{hh\rightarrow\phi\phi}\lesssim H$ before the EWPT, and since $\Gamma_{hh\rightarrow\phi\phi}\propto T$ and $H\propto T^2$, the strongest constraint is at $T_{EW}$. This yields an upper bound on the Higgs-dark scalar field coupling:
\begin{eqnarray}
g\lesssim 8\times10^{-4}\,\left(\frac{g_{*}}{100}\right)^{1/8}~.\label{g bound evaporation}
\end{eqnarray}
Another possibility for the condensate's evaporation is the perturbative production of $\phi$ particles from background field oscillations. For $T>T_{EW}$, $\phi$-particles are effectively massless and interact with the background field through the coupling $\mathcal{L}_{int}=-\frac{3}{2}\,\lambda_{\phi}\,\phi^{2}\delta\phi^{2}$, which can be obtained by decomposing the field into a background component and particle fluctuations $\delta \phi$. The process of particle production from an oscillating background field with a quartic potential has been studied in detail in Refs. \cite{Kainulainen:2016vzv,Greene:1997fu,Ichikawa:2008ne}, yielding a particle production rate
\begin{eqnarray}
\Gamma_{\phi\rightarrow\delta\phi\delta\phi} & \simeq & 4\times10^{-2}\,\lambda_{\phi}^{3/2}\,\phi~,\label{decay phi_0^4}
\end{eqnarray}
which is valid above $T_{EW}$, whereas after the EWPT $\phi$ particles gain a mass and the process becomes kinematically forbidden. Since $\Gamma_{\phi\rightarrow\delta\phi\delta\phi}\propto T$ in the quartic oscillations regime, we again have that the strongest constraint is at $T_{EW}$ where $\phi\simeq \phi_{EW}$, yielding an upper bound on the dark scalar self-coupling:
\begin{equation}
\lambda_{\phi}<6\times10^{-10}\left(\frac{g_{*}}{100}\right)^{1/5}\left(\frac{r}{0.01}\right)^{-1/5}\xi^{1/10}.\label{bound on lambda from cond evaporation}
\end{equation}
If Eqs. (\ref{g bound evaporation}) and (\ref{bound on lambda from cond evaporation}) are satisfied, the dark scalar never thermalizes with the cosmic plasma and behaves effectively as an oscillating condensate throughout its whole cosmic history. Given the relation between couplings obtained in Eq.~(\ref{g and lambd}), we see that Eq.~(\ref{bound on lambda from cond evaporation}) gives the strongest constraint, limiting the viable DM mass range to $\lesssim 1$ MeV. Our DM candidate must thus be a light particle weakly coupled to the Higgs boson, but due to its condensate nature it nevertheless behaves as cold dark matter after the EWPT.

This weak coupling to the Higgs field has nevertheless quite significant implications, since at the minimum the $\phi$ and $h$ scalars exhibit a small mass mixing, with mixing parameter $\epsilon=g^2\phi_0\mathrm{v}/ m_h^2$, which can be written as: 
\begin{equation}
\epsilon\simeq4\times10^{-13}\,\left(\frac{m_{\phi}}{7\,\mathrm{keV}}\right)\,\left(\frac{0.5}{x_{DM}}\right).\label{eff vertex DM final}
\end{equation}
Equivalently, the physical mass eigenstates are a small admixture of the original $\phi$ and $h$ fields. This implies that the dark scalar can decay into the same decay channels as the Higgs boson, provided that they are kinematically accessible, but with a decay width suppressed by $\epsilon^2$ w.r.t.~the corresponding Higgs partial width. With the DM mass bound obtained above, the only kinematically accessible decay channel is into photon pairs \cite{foot}, with decay width $\Gamma_{\phi\rightarrow\gamma\gamma}=\epsilon^2\Gamma_{H^{*}\rightarrow\gamma\gamma}$, where $H^{*}$ represents a virtual Higgs state with invariant mass $p^2=m_\phi^2$. The partial decay width of a virtual Higgs into photons is given by \cite{Djouadi:2005gi}:
\begin{equation} \label{higgs decay into photons}
\Gamma_{H^{*}\rightarrow\gamma\gamma}=\frac{G_{F}\,\alpha_{QED}^{2}\,m_{\phi}^{3}}{128\sqrt{2}\,\pi^{3}}F^2~,
\end{equation}
where, $G_{F}=1.17\times10^{-5}\,\mathrm{GeV}^{-2}$ is Fermi's  constant, $\alpha_{QED}\simeq1/137$ is the fine structure constant and
\begin{equation} \label{F_factor}
F= \bigg|\sum_{f}\,N_{c}\,Q_{f}^{2}\,A_{1/2}^{H}\left(\tau_{f}\right)+A_{1}^{H}\left(\tau_{\mathbf{w}}\right)\bigg|\simeq {11\over 3}
\end{equation}
accounts for the loop contributions of all charged fermions and the $W$ boson to the decay, with $\tau_i= 4m_i^2/m_\phi^2\gg 1$ for all particle species involved. We note that for the decay of a virtual Higgs all charged fermions give essentially the same contribution, whereas for an on-shell Higgs boson only the top quark contributes significantly. We then obtain for the DM lifetime:
\begin{equation} \label{decay time}
\tau_{\phi}\simeq7\times10^{27}\left(\frac{7\,\mathrm{keV}}{m_{\phi}}\right)^{5}\,\left(\frac{x_{DM}}{0.5}\right)^{2}\,\mathrm{sec},
\end{equation}
which is much larger than the age of the Universe, but can nevertheless lead to an observable monochromatic line in the spectrum of galaxies and galaxy clusters.

In fact, the XMM-Newton X-ray observatory has recently discovered a line at 3.5 keV, which is present not only in the Galactic Center (GC) but also in other astrophysical systems such as Andromeda and the Perseus cluster \cite{Bulbul:2014sua,Boyarsky:2014jta,Boyarsky:2014ska,Cappelluti:2017ywp}. The origin of this line has led to several interesting proposals in the literature, in particular the possibility of it resulting from DM decay or annihilation \cite{Higaki:2014zua,Jaeckel:2014qea,Dudas:2014ixa,Queiroz:2014yna,Cappelluti:2017ywp,Heeck:2017xbu}. Although other astrophysical processes have been considered \cite{Jeltema:2014qfa}, there are also some independent studies that contest them \cite{Boyarsky:2014paa,Bulbul:2014ala,Iakubovskyi:2015kwa}. There is still an ongoing controversy regarding the discovery of this line in dwarf galaxies, such as Draco. While some groups indicate that this line is not present in such objects \cite{Jeltema:2015mee}, others claim that the line is there but is too faint to be observed with current technology. The authors of \cite{Ruchayskiy:2015onc}, in particular, conclude that observations of dwarf spheroidal galaxies cannot exclude the DM decay explanation of the line.

\begin{figure}[h]
\begin{centering}
\includegraphics[scale=0.40]{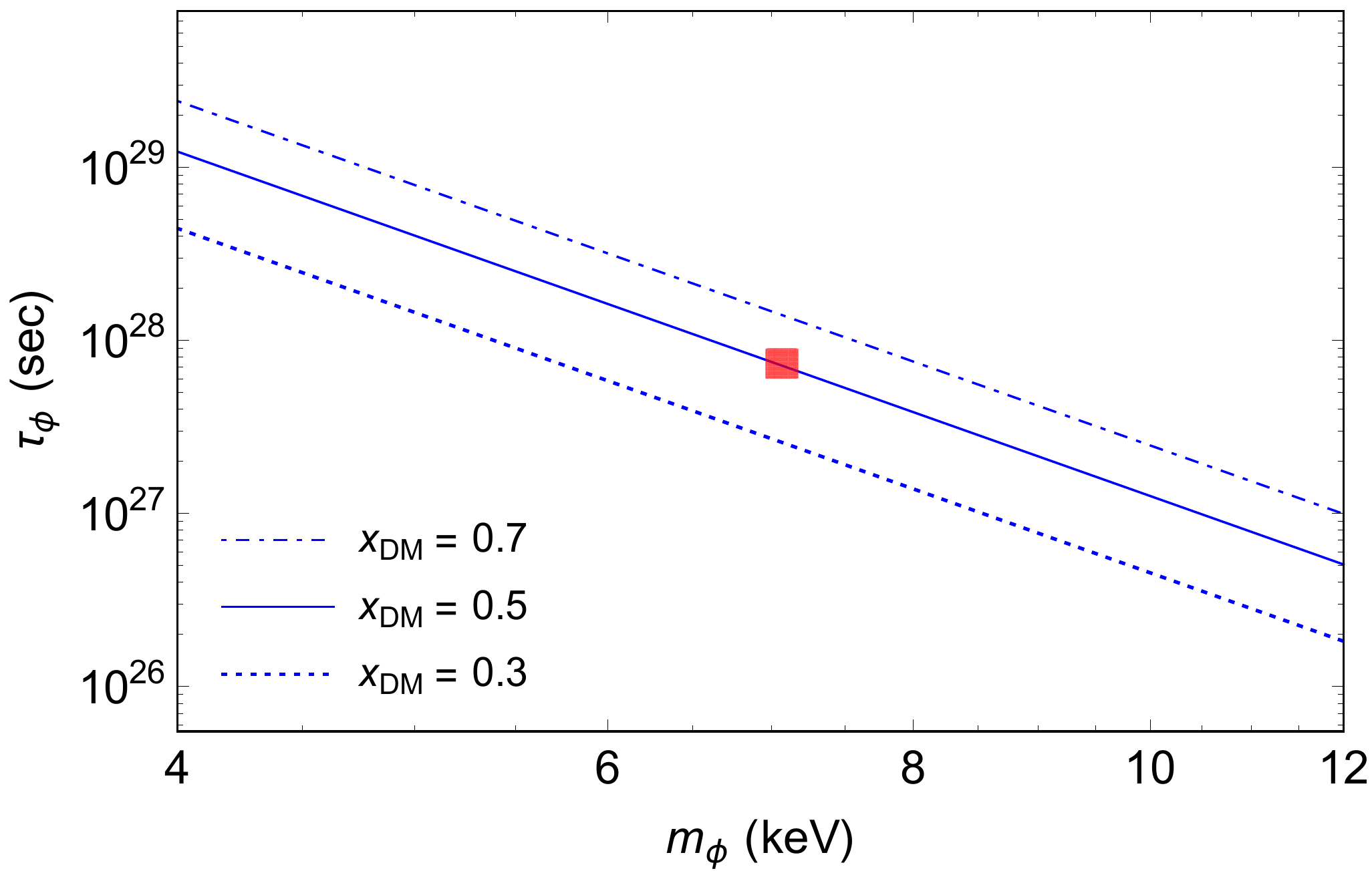}
\par\end{centering}
\vspace{-0.2cm}
\caption{Lifetime of the scalar field dark matter as a function of its mass, for different values of $x_{DM}\lesssim 1$ parametrizing the uncertainty in the value of the field oscillation amplitude after the EWPT. The shaded horizontal band corresponds to the values of $\tau_\phi$ that can account for the intensity of the 3.5 keV X-ray line observed by XMM-Newton for a mass around $7\,\,\mathrm{keV}$ including the uncertainty in the photon energy combining different observations \cite{Ruchayskiy:2015onc}.}
\label{plot}
\end{figure}

The analysis in Refs. \cite{Boyarsky:2014ska,Ruchayskiy:2015onc} has shown that the intensity of the line observed in the GC, Andromeda and the Perseus cluster could be explained by the decay of a DM particle with a mass of $\simeq$ 7 keV and a lifetime in the range $\tau_{\phi}\sim\left(6-9\right)\times10^{27}$ sec. This would also explain the absence of such a line in the blank-sky data set. In the case of our dark scalar field model, setting the field mass to this value, we {\it predict} a DM lifetime exactly in this range, up to some uncertainty in the value of the field amplitude after the EWPT parametrized by $x_{DM}\lesssim 1$. This is illustrated in Fig.~\ref{plot}.  

\vspace{-0.1cm}

For this mass value, we have $g\simeq3\times10^{-8}$ and, from Eq. (\ref{g and lambd}), $\lambda_{\phi}\simeq4\times10^{-20}$, which satisfy the constraints
in Eqs. (\ref{g bound evaporation}) and (\ref{bound on lambda from cond evaporation}). Note that such a small quartic self-coupling for the scalar field is technically natural, since quantum corrections to this coupling from interactions with the Higgs field are $\lesssim g^4\sim 10^{-30}$.

We do not aim to explain here the smallness of these couplings, which would require going beyond the effective theory approach that we have followed in this work. We nevertheless note that small couplings can be naturally obtained in the context of extra-dimensional geometries, as in the warped dark scalar field scenario developed in Ref. \cite{Bertolami:2016ywc}.  This issue is also essentially on the same footing as explaining the smallness of fermion masses, namely the electron.

The most impressive feature of our model is that, although it originally
involves four parameters - the couplings $g$ and $\lambda_{\phi}$,
the non-minimal coupling $\xi$ and the scale of inflation $r$, the
last two do not affect the predictions of the model regarding the
3.5 keV line. The role of $\xi$ is simply to suppress potential CDM
isocurvature perturbations, while $r$ only sets the field amplitude
at the onset of the radiation era. Since after the EWPT the field
starts oscillating about the value $\phi_{0}$ (depending on g and
$\lambda_{\phi}$) and its amplitude is also of this order, the present
DM abundance, the field mass and its decay width are just dependent
on the couplings $g$ and $\lambda_{\phi}$, yielding three observable
quantities determined by only two parameters. In other words, the
field loses the memory of its initial conditions at the EWPT, and
$\xi$ and $r$ do not affect its dynamics afterwards. Therefore,
if the scalar field accounts for all the DM in the Universe, its mass
and decay width are effectively only dependent on a single parameter.

The smallness of $g$ may make it hard to probe the dark scalar interpretation of the 3.5 keV X-ray line in the laboratory. An obvious possibility is to look for invisible Higgs decays, but the predicted branching ratio for $H\rightarrow \phi \phi$ is $\sim 10^{-27}$, which is unrealistic to probe in the near future. Mass mixing also gives rise to Higgs-dark scalar oscillations, but again with a small probability suppressed by $\epsilon^2$. However, if astrophysical observations are able to exclude other explanations for the 3.5 keV line or even clearly confirm a correlation between the line's intensity and the cosmic DM distribution, this should serve as motivation for extremely precise measurements of the Higgs properties in the future. 

The dark scalar coupling to photons, of the form $\phi F_{\mu\nu}F^{\mu\nu}$ due to the $\phi-h$ mixing, could also be used to look for X-ray photon-DM conversion in an external electric or magnetic field in light-shining-through-a-wall experiments akin to those looking for axion-like particles, for which conversion probabilities are also very small \cite{Redondo:2010dp}.

The scenario proposed in this Letter may also be of interest for different values of $m_\phi$ if the DM interpretation of the 3.5 keV line is refuted. In particular, for $m_\phi\lesssim 0.1$ eV, the scalar field will exhibit a coherent behavior on galactic scales, and its mixing with photons and other SM particles through the Higgs portal may lead to small oscillations of fundamental constants, namely $\alpha_{QED}$ and the electron mass. There are already proposals for detecting similar oscillations using mass-resonant detectors \cite{Arvanitaki:2014faa,Arvanitaki:2015iga,Arvanitaki:2016fyj,Stadnik:2014tta,Stadnik:2015kia,Stadnik:2015xbn,Hees:2016gop,Stadnik:2016zkf}, and we will explore this possibility in more detail in a companion paper. 

In our analysis, we have assumed that the Lagrangian (\ref{Lagrangian}) exhibits a U(1) symmetry. If the coupling between the Higgs and the dark scalar has  a negative sign, the symmetry is spontaneously broken at the EWPT. The consequences of this symmetry breaking depend on whether it is a global U(1) symmetry or a gauged U(1) symmetry. On one hand, in the former case, $\phi$ may decay into massless Goldstone bosons, and may survive until the present day only for $\lambda_{\phi}<2\times10^{-32}\,\left(\frac{x_{DM}}{0.5}\right)^{2/5}$. This limits the viable range for the dark matter mass to $m_{\phi}\lesssim5\,\mathrm{eV}$. Therefore, this scenario cannot explain the 3.5 keV mass, although it still allows for a lighter dark matter candidate. On the other hand, we may consider a spontaneously broken U(1) gauge symmetry, where the Goldstone boson is absorbed into the longitudinal component of the massive gauge boson. If the gauge boson acquires a sufficiently large mass, $\phi$ decay will be kinematically blocked, which imposes only a mild constraint on the gauge coupling, $e'>\sqrt{2\lambda_{\phi}}$, noting that the quartic self-coupling is typically very small. The dark scalar's oscillations may induce gauge boson production above the Electroweak scale, similarly to the case where $\phi$ particles are produced perturbatively by the background scalar field oscillations. Nonetheless, after the EWPT this process becomes kinematically forbidden, and to prevent a significant production of gauge bosons we may impose an upper bound on the square of the gauge coupling of the order of the limit on $\lambda_\phi$ (see Eq. (\ref{bound on lambda from cond evaporation})), and which may be thus compatible with the above-mentioned lower bound. The  “dark photons” could, in addition, be thermally produced in the early Universe in the presence of kinetic mixing with ordinary photons, but since there are no particles charged under both U(1) gauge groups, such mixing is absent in our model. In fact, $2\leftrightarrow 2$ scattering processes involving dark and visible photons are only generated through the Higgs-portal scalar mixing, which yields a dimension-6 operator that is suppressed with respect to the dark scalar’s effective (visible) photon coupling by the smallness of the dark U(1) gauge coupling. Hence, within the parametric regime described above, the dark photons are not significantly produced in the early Universe and can neither make a significant contribution to the dark matter abundance nor lead to the condensate’s decay or evaporation. A more detailed study of the cosmological implications of the spontaneous symmetry breaking for this model is done in Ref. \cite{Cosme:2018nly}.

Our model has also other interesting phenomenological consequences, for instance the formation of cosmic strings due to the U(1) symmetry breaking at the EWPT. The energy density of cosmic strings in the scaling regime, $\rho_{s}\sim\frac{\mu}{t^{2}}$, where
$\mu$ is the string's energy per unit length \cite{Vachaspati:2000cq}, follows the background density $\rho_{c}\sim\frac{1}{G\,t^{2}}$. Their ratio $\rho_s/\rho_c\sim G\mu\simeq10^{-6}\left(\phi_0/10^{16}\,\mathrm{GeV}\right)^{2}$ \cite{Kolb:1990vq} is, however, extremely small for the values of interest in our model, where $\phi_0\sim 26$ TeV. We also note that our model is viable and provides the same dynamics and predictions if, instead of a complex scalar field with a U(1) gauge symmetry, we consider a real scalar field with a $\mathbb{Z}_{2}$ symmetry. Although the $\mathbb{Z}_{2}$ spontaneous symmetry breaking leads to the formation of a network of domain walls at the EWPT, this network may decay if there is a bias in the initial configuration of the field towards one of the potential minima, which could likely result from field fluctuations during inflation. This possibility was first studied in Ref. \cite{Larsson:1996sp} and can be applied to our model, since the dark scalar field is never in thermal equilibrium with the cosmic plasma and so the bias induced by inflation can last until the EWPT and therefore wipe out the domain wall network generated by the symmetry breaking before it modifies the cosmic evolution \cite{Cosme:2018nly}.

In summary, we have shown, for the first time, that an oscillating scalar field coupled to the Higgs boson is a viable DM candidate that can explain the observed 3.5 keV X-ray line. The simplicity of our model, based on the assumed scale-invariance of DM interactions, makes it extremely predictive, with effectively only a single free parameter upon fixing the present DM abundance. Hence, our scenario {\it predicts} a 3.5 keV X-ray line with the observed properties for the corresponding value of the DM mass.

\vspace{1.5cm} 

\vspace{0.5cm} 

\noindent
{ \bf Acknowledgements}

\noindent


C.\,C. is supported by the Funda\c{c}\~{a}o para a Ci\^{e}ncia e Tecnologia (FCT) grant PD/BD/114453/2016. J.\,G.\,R. is supported by the FCT Investigator Grant No.~IF/01597/2015 and partially by the H2020-MSCA-RISE-2015 Grant No. StronGrHEP-690904 and by the CIDMA Project No.~UID/MAT/04106/2013.

 \vfill


\begin{thebibliography}{9}
\bibitem{foot2}
Note that $T_{EW}$ is typically below the critical temperature
for the electroweak crossover by a factor $\sim2$ and that
the potential minimum smoothly evolves from the origin
towards $\phi_{0}$. This introduces an $\mathcal{O}\left(1\right)$ uncertainty in the
amplitude of field oscillations at $T_{EW}$ that translates into
a similar uncertainty in the $x_{DM}$ parameter.
\bibitem{foot}
The decay $\phi\rightarrow e^+e^-$ may be possible for DM masses saturating the obtained bound, but in this case condensate evaporation may not be negligible and, hence, we do not consider this possibility in detail.
\end{thebibliography}
\end{document}